\documentstyle[prb,aps,multicol,eqsecnum]{revtex}

\renewcommand{\narrowtext}{\begin{multicols}{2}}
\renewcommand{\widetext}{\end{multicols}}

\begin{document}
\title{Spin and orbital effects in a $2D$ electron gas in a random magnetic
field.}
\author{K.B. Efetov${}^{1,2}$, V. R. Kogan$^{1,2}$ }
\address{${}^1$ Theoretische Physik III, \\
Ruhr-Universit\"at Bochum, 44780 Bochum, Germany\\
${}^2$ L. D. Landau Institute for Theoretical Physics, 117940 Moscow, Russia}
\date{\today}
\maketitle
\draft

\begin{abstract}
Using the method of {\it superbosonization} we consider a model of a random
magnetic field (RMF) acting on both orbital motion and spin of electrons in
two dimensions. The method is based on exact integration over one particle
degrees of freedom and reduction of the problem to a functional integral
over supermatrices $Q({\bf r},{\bf r^{\prime }})$. We consider a general
case when both the direction of the RMF and the $g$-factor of the Zeeman
splitting are arbitrary. Integrating out fast variations of $Q$ we come to a
standard collisional unitary non-linear $\sigma $-model. The collision term
consists of orbital, spin and effective spin-orbital parts. For a particular
problem of a fixed direction of RMF, we show that additional soft
excitations identified with spin modes should appear. Considering $\delta $%
-correlated weak RMF and putting $g=2$ we find the transport time $\tau
_{tr} $. This time is $2$ times smaller than that for spinless particles.
\end{abstract}


\narrowtext

\section{Introduction}

\label{introduction}

Models of a random magnetic field (RMF) acting on electrons in two
dimensions are intensively studied in mesoscopic physics. There are direct
experiments on high-mobility heterostructures subjected to a magnetic field
of randomly pinned flux vortices in a type-II superconducting gate \cite%
{GBG92}, type-I superconducting grains \cite{S94} or a demagnetized
ferromagnet \cite{M95}. For theoreticians, the RMF models are important as
an example of systems with an interaction reduced to a gauge field. These
models arise in theory of composite fermions in the fractional quantum Hall
effect near half-filling\cite{HLR93} as well as in a description of doped
Mott insulators\cite{IL89}.

From the theoretical point of view, one of the most interesting tasks in the
study of any model with a disorder is to determine the large scale behavior
of electrons and find universal properties of the model. In doing so, one
may come either to a metal or insulating behavior. In the latter case
electron wave finctions should be localized. In the context of the RMF
models the localization have been discussed in many numerical works, which
resulted in three different conclusions: a) localization of all the states %
\cite{SN93}$^{-}$\cite{BSK98}; b) existence of a band of delocalized states %
\cite{KWAZ93}$^{-}$\cite{SW00}; c) localization of all the states except
those in the band center\cite{MW96}$^{-}$ \cite{C01}.

Analytically, the RMF models were studied by the standard diagrammatic
technique\cite{AMW94} as well as using different non-linear $\sigma $-models %
\cite{AMW94}$^{-}$\cite{EKrmf03}. The final conclusion drawn by using the methods
was that the RMF models belonged to the usual unitary class of universality
with localization in two dimensions (provided the correlations of the RMF 
$\langle B_{q}B_{-q}\rangle $ at small $q\rightarrow 0$ increase slower than 
$1/q^{2}$ , Ref.\cite{ETS01}).

Ususally, when deriving the proper $\sigma $-model for the RMF problems one
uses the standard scheme\cite{E83} based on the saddle-point approximation.
Calculations presented in Refs. \cite{AMW94}$^{-}$\cite{ETS01} are performed in
this way. From the point of view of the conventional perturbation theory
this approximation corresponds to the self-consistent Born approximation
(SCBA), which is not good for a long range disorder. The same approximation
has been used in the diagrammatic approach of Ref.\cite{AMW94}.

In addition to the difficulty of a description of the long range disorder,
the standard scheme is not convenient for a generalization of the RMF model
for the case when spin degrees of freedom are important. This is because the
effect of the magnetic field on the orbital electron motion is accounted for
by adding a vector potential in the Hamiltonian, whereas the interaction
with the electron spin is described by the magnetic field itself. As the
correlations of the vector potentials and the magnetic fields are different,
it is not easy to consider both the effects on equal footing.

Partly this problem has been resolved in a recent paper \cite{ET02} for free
electrons in $2D$ with the $g$-factor $g=2$ and a magnetic field
perpendicular to the plane. Unfortunately, a mathematical trick of replacing
the initial Hamiltonian by a Dirac Hamiltonian used in that paper, which was
the basis of the suggested calculation scheme, can be applied only for this
particular system.

Another analytical method suggested in Ref.\cite{EKrmf03} enables one to
avoid using the saddle-point approximation and to obtain a ballistic 
$\sigma$- model applicable at all distances down to the Fermi wavelength 
$\lambda_{F}$. This method is based on using quasiclassical equations of motions
which contain not the vector potential but the magnetic field only. As the 
vector potential itself does not enter the ballsitic $\sigma $-model one can 
include rather easily the Zeeman term without extra assumptions about the value 
of the $g$-factor and the direction of the magnetic field. However, this is still 
not the most general method because it is essentially based on the quasiclassical 
approximation and therefore short range correlations of the magnetic field cannot 
be considered.

In the present paper we use a new method of {\it superbosonization}
suggested recently\cite{EST04}. The method is based on the exact integration
over the one particle motion and reformulation of the initial fermionic
problem in terms of a functional integral over supermatrices $Q({\bf r},{\bf %
r^{\prime }})$. As it has been shown in Ref. \cite{EST04}, in the
quasiclassical regime (smooth disorder and lengths larger than $\lambda _{F}$%
) the matrix $Q({\bf r},{\bf r^{\prime }})$ corresponds to the matrix $Q_{%
{\bf n}}({\bf r})$ of Ref.\cite{EK03}. The method of the superbosonization
has several advantages. First, it uses neither saddle-point nor
quasiclassical approximations and is exact. Both short and long range
disorder can be considered on its basis. Second, integration over one
particle motion is carried out before disorder averaging. The latter enables
one to consider interaction effects related to the random scattering more
carefully. Finally, the method results in a simple expression for the energy
in terms of the matrices $Q({\bf r},{\bf r^{\prime }})$ and makes the
disorder averaging a rather simple procedure.

The new theory is invariant with respect to rotations of the matrix $Q({\bf r%
},{\bf r^{\prime }})$ in the superspace provided they commute with the
Hamiltonian $\hat{H}$. This makes possible separation of the massive modes
from the soft ones and we can integrate them out. We show that this
procedure is justified only in the regime when the scattering effects do not
result in a strong coupling between electrons. For the short range disorder
this is so over distances exceeding the correlation length of the disorder (or 
the Fermi wavelength $\lambda _{F}$) whereas for the smooth disorder the 
coupling becomes sufficiently weak beyond the {\it Lyapunov length}. The 
latter case was discussed in many details in Refs. \cite{AL96}$^{,}$\cite{GM02}.

The low energy theory found is described by a ballistic nonlinear $\sigma $-
model with a collision term. This term consists of three parts that can be
related to the orbital, spin and some effective spin-orbital scattering.
Similarly to the model with magnetic impurities (see Ref.\cite{E83}), the
second scattering results generally in the relaxation of all spin modes. At
the same time, we show that in the model with a fixed direction of RMF an
additional soft mode corresponding to fluctuations of the spin along the
field should appear. Finally, using standard method of integrating out the
angle modes we come to the conventional unitary diffusive $\sigma $- model
and find the transport time $\tau _{tr}$.

The paper is organized as follows. In Sec.\ref{bosonization} we discuss the
superbosonization procedure and give an alternative derivation of
superbosonized theory. At the end of the section we discuss symmetry
properties of the obtained model and derivartion of the non-linear $\sigma $%
- model.

In Sec.\ref{collision} we introduce the main definitions of RMF model
involved, derive the $\sigma $- model valid in the collisional regime and
discuss conditions of its applicability.

In Sec.\ref{diffusion} the same problem is considered in the diffusive
limit. We calculate with the help of the obtained diffusive $\sigma $-model
the transport time and the spin susceptibility.

\section{Superbosonization and derivation of the $\protect\sigma $-model.}

\label{bosonization}

Below we consider a two-dimensional ($2D$) electron gas placed in a static
inhomogeneous magnetic field. The field is assumed to act both on the
orbital motion and electron spin. Our consideration is based on the new
method of {\it superbosonization} proposed in recent paper \cite{EST04}.
This method uses exact integration over electron degrees of freedoms and
reformulation of the problem in terms of integrals over supermatrices with a
rotational symmetry (nonlinear $\sigma $-model).

Before starting the study of RMF problem we would like to give alternative
derivation of the superbosonized model. Although, the scheme of the
derivation presented in Ref. \cite{EST04} is straightforward and exact, the
final representation of the Green functions in terms of a functional
integral over the supermatrices can be obtained even in a more simple way.
Following the standard way (see the book \cite{E83}) we introduce first a
generating functional $Z[a]$: 
\begin{equation}
Z[a]=\int \exp \bigl(-L_{a}[\psi ]\bigr)D\psi  \label{e1.1}
\end{equation}%
where the Lagrangian $L_{a}[\psi ]$ has the form 
\begin{eqnarray}
L_{a}[\psi ] &=&-i\int \bar{\psi}({\bf r})\left( \hat{H}_{{\bf r}}-\epsilon +%
\frac{\omega }{2}+\frac{\omega +i\delta }{2}\Lambda \right) \psi ({\bf r})d%
{\bf r}+  \nonumber \\
&&i\int \bar{\psi}({\bf r})a({\bf r},{\bf r^{\prime }})\psi ({\bf r^{\prime }%
})d{\bf r}d{\bf r^{\prime }},  \label{e1.2}
\end{eqnarray}%
$\hat{H}_{{\bf r}}$ is Hamiltonian of the initial model (we write it below)
and $\psi({\bf r})$ is a supervector. The conjugated supervector $\bar{\psi}({\bf r})$ 
is related to $\psi(\bf r)$ according to the following definition:
\begin{equation}
\bar{\psi}({\bf r})=(C\psi)^T({\bf r})
\label{e1.2a}\end{equation}
The structure of the supervectors $\psi({\bf r})$, $\bar{\psi}({\bf r})$ as well 
as the form of the matrix $C$ depends on the problem involved (for details see 
Ref.\cite{E83}). For the problem of electrons with spin considered below 
$\psi({\bf r})$ should be a $16$-component supervector.

The Lagrangian $L_{a}[\psi ]$, Eq.(\ref{e1.2}), differs from the
conventional one by the presence of the source $a({\bf r},{\bf r^{\prime }})$. 
In general, the source term is for the problem considered an arbitrary $16\times 16$ 
supermatrix depending on two coordinates ${\bf r}$, ${\bf r^{\prime }}$. 
Differentiating in elements of this matrix one can obtain various correlation 
functions. For example, the level-level correlation function $R(\omega )$ can 
be written as 
\begin{equation}
R(\omega )=\frac{1}{2}-\frac{1}{2(\pi \nu V)^{2}}\lim_{\alpha _{1},\alpha
_{2}=0}{\rm Re}\frac{\partial ^{2}}{\partial \alpha _{1}\partial \alpha _{2}}%
Z[a]  \label{e1.3}
\end{equation}%
provided the source $a({\bf r},{\bf r^{\prime }})$ is taken in the following
form 
\[
a({\bf r},{\bf r'})=\hat{a}({\bf r})\delta({\bf r}-{\bf r'})\]
\begin{equation}
\hat{a}({\bf r})=\left( 
\begin{array}{ccc}
\hat{\alpha}_{1} & 0 &  \\ 
0 & -\hat{\alpha}_{2} & 
\end{array}%
\right) ,\text{ \ \ }\hat{\alpha}_{1,2}=\frac{\alpha _{1,2}}{2}(1-k),
\label{e1.4}
\end{equation}%
$k=\pm 1$ in fermionic and bosonic blocks respectively. As supermatrix 
$\psi_{\alpha}({\bf r})\bar{\psi}_{\beta}({\bf r'})$ is self-conjugated 
one may consider only self-conjugated sources:
\begin{equation}
\bar{a}({\bf r},{\bf r'})\equiv Ca^T({\bf r'},{\bf r})C^T=a({\bf r},{\bf r'})
\label{e1.4a}\end{equation}
Sources with constraint Eq.(\ref{e1.4a}) have minimal number of elements 
required when finding an arbitrary correlation function.  

The integral defined in Eqs.(\ref{e1.1}), ({\ref{e1.2}}) is gaussian and can
be readily calculated: 
\begin{equation}
Z[a]=\exp \left( \frac{1}{2}{\rm Str}\ln \left[ \hat{H}-\epsilon +\frac{%
\omega }{2}+\frac{\omega +i\delta }{2}\Lambda -a\right] \right)  \label{e1.5}
\end{equation}%
The logarithmic derivative of the partition function $Z[a]$, Eq.({\ref{e1.5}}%
), in $a$ is a matrix Green function $G({\bf r},{\bf r^{\prime }})$: 
\begin{equation}
\frac{\delta \ln Z[a]}{\delta a({\bf r},{\bf r^{\prime }})}=\frac{i}{2}G(%
{\bf r},{\bf r^{\prime }})  \label{e1.6}
\end{equation}%
satisfying the following equation 
\begin{equation}
\left( \hat{H}_{{\bf r}}-\epsilon +\frac{\omega }{2}+\frac{\omega +i\delta }{2}\Lambda
-a\right) G({\bf r},{\bf r^{\prime }})=i\delta \left( {\bf r-r}^{\prime
}\right)  \label{e1.7}
\end{equation}%
Calculating the Green function $G({\bf r},{\bf r^{\prime }})$ from Eq. (\ref%
{e1.7}) and using Eq.(\ref{e1.6}) one can find the partition function $Z[a]$%
, Eqs.(\ref{e1.1}),(\ref{e1.5}).

What we want to show now is that the Green function, Eq. (\ref{e1.7}), can
be represented exactly as an integral over supermatrices $Q\left( 
{\bf r,r}^{\prime }\right) $ in the following form 
\begin{equation}
G({\bf r},{\bf r^{\prime }})=\tilde{Z}^{-1}[a]\int Q({\bf r},{\bf r^{\prime }%
})\exp \bigl(-F_{a}[Q]\bigr)DQ  \label{e1.8}
\end{equation}%
where $\tilde{Z}[a]$ is a new partition function 
\begin{equation}
\tilde{Z}[a]=\int \exp \bigl(-F_{a}[Q]\bigr)DQ  \label{e1.9}
\end{equation}%
and the functional $F_{a}[Q]$ has the form 
\begin{eqnarray}
F_{a}[Q]=\frac{i}{2}{\rm Str}\int  &&\left( \hat{H}_{{\bf r}}-\epsilon +%
\frac{\omega }{2}+\frac{\omega +i\delta }{2}\Lambda \right) \times  
\nonumber \\
&&\delta ({\bf r}-{\bf r^{\prime }})Q({\bf r},{\bf r^{\prime }})d{\bf r}d%
{\bf r^{\prime }}+  \label{e1.10}
\end{eqnarray}%
\[
\frac{1}{2}{\rm Str}\ln Q-\frac{i}{2}{\rm Str}\int a({\bf r},{\bf r^{\prime }%
})Q({\bf r^{\prime }},{\bf r})d{\bf r}d{\bf r^{\prime }}
\]
Supermatrix $Q({\bf r},{\bf r'})$ in the integral Eq.(\ref{e1.8}) is not 
arbitrary and should have a certain structure. First, it must be self-conjugated:
\begin{equation}
\bar{Q}({\bf r},{\bf r^{\prime }})=Q({\bf r},{\bf r^{\prime }})  \label{e1.11}
\end{equation}
to provide the same for the result of the integration in Eq.(\ref{e1.8}).  

Second, the structure of the supermatrix $Q({\bf r},{\bf r'})$ should be defined 
such that the energy $F_a[Q]$, Eq.(\ref{e1.10}), would have a minimum. One can show 
that this takes place if ${\rm Str}(Q^2)$ is non-negative. The latter requirement 
is fulfilled for the supermatrices with the following constraint:
\begin{equation}
Q_{\perp}({\bf r},{\bf r'})=KQ_{\perp}^+({\bf r'},{\bf r})K,\;\;\;
K=\left(
\begin{array}{ccc}
1 & 0 \\
0 & k
\end{array}\right),
\label{e1.10a}\end{equation}
where $Q_{\perp}=(1/2)\Lambda[\Lambda,Q]$, $k$ is the same as in Eq.(\ref{e1.4}). 
Relation (\ref{e1.10a}) will be used below when formulating the structure of a 
$\sigma$ model.
 
Eqs. (\ref{e1.8}-\ref{e1.10}) have been suggested in Ref.\cite{EST04} and
we want to demonstrate now a simpler way of their derivation. In order to
prove Eq.(\ref{e1.8}) we write the following identity 
\[
-2i\tilde{Z}^{-1}[a]\int \biggl[\int \frac{\delta \exp \left( -\frac{1}{2}%
{\rm Str\ln Q}\right) }{\delta Q({\bf r^{\prime \prime }},{\bf r})}Q({\bf %
r^{\prime \prime }},{\bf r^{\prime }})d{\bf r^{\prime \prime }}\biggr]\times 
\]
\[
\exp \left( -\frac{i}{2}{\rm Str}\left[ \hat{H}-\epsilon +\frac{\omega }{2}%
+\frac{\omega +i\delta }{2}\Lambda -a\right] Q\right) DQ=  \]
\begin{equation}
=i\delta ({\bf r}-{\bf r^{\prime }}),  \label{e1.12}
\end{equation}%
and integrate over $Q$ by parts. The derivative $\delta /\delta Q$ should
act now on both $Q$ and the exponential. At this point, the supersymmetry
plays a crucial role. Differentiating first the supermatrix $Q$ we obtain
the supermatrix product $\left( \delta /\delta Q\right) Q.$ As the number of
the anticommuting variables in the sum over the matrix elements is equal to
the number of the boson ones and the derivatives have the oposite signs,
this matrix product vanishes. Differentiating the exponential only we come
to the following equation:    
\[
\tilde{Z}^{-1}[a]\int d{\bf r^{\prime \prime }}\left( \hat{H}
-\epsilon +\frac{\omega }{2}+\frac{\omega +i\delta }{2}\Lambda -a\right) (%
{\bf r},{\bf r^{\prime \prime }})\times   \]
\begin{equation}
\int Q({\bf r^{\prime \prime }},{\bf r^{\prime }})\exp \bigl(-F_{a}[Q]%
\bigr)DQ=i\delta ({\bf r}-{\bf r^{\prime }})  \label{e1.13}
\end{equation}
Eq. (\ref{e1.13}) proves immediately that the integral Eq.(\ref{e1.8}) does
satisfy Eq.(\ref{e1.7}) and we have really the alternative representation of
the Green function in terms of an integral over the supermatrices $Q$. 

Integrating Eq.(\ref{e1.8}) over the source $a({\bf r},{\bf r^{\prime }})$
we conclude that partition functions $Z[a]$, Eqs.(\ref{e1.1}, \ref{e1.5})
and $\tilde{Z}[a]$, Eq.(\ref{e1.9}), are proportional to each other with a
coefficient that is independent of the source. Putting $a({\bf r},{\bf %
r^{\prime }})=0$ in the definitions of the both functions and finding that $%
Z[a=0]=\tilde{Z}[a=0]=1$ due to supersymmetry we come to the equality: 
\begin{equation}
Z[a]=\tilde{Z}[a]  \label{e1.14}
\end{equation}%
We emphasize that the relation (\ref{e1.14}) is exact and does not depend on
the form of the Hamiltonian $\hat{H}_{{\bf r}},$ Eq.(\ref{e1.2}), and the source $a(%
{\bf r},{\bf r^{\prime }})$ provided both the partition functions are
defined in Eqs.(\ref{e1.1}), (\ref{e1.9}) in terms of the superintegrals.

In the low energy limit, the field theory specified by Eq.(\ref{e1.10}) can
be reduced to an effective nonlinear $\sigma $-model\cite{EST04}. A
possibility of this reduction is related to the invariance of the free
energy functional $F_{a}$ with respect to the rotations of the supermatrix $Q$ 
\begin{equation}
Q\rightarrow VQ\bar{V}  \label{e1.15}
\end{equation}%
for the case when the unitary matrix $V\bar{V}=1$ commutes with the
Hamiltonian $\hat{H}$, Eq.(\ref{e1.2}). This means that the low energy limit
of the model is determined by rotations $V$ with small values of commutator 
$[\hat{H},V]$. For different problems the smallness of the commutator can be
provided by imposing on the rotations $V$ different conditions. For RMF
model we discuss this point at the end of the section \ref{collision}.

In order to reduce the general formula, Eq.(\ref{e1.10}), to a $\sigma $-
model containing only the rotations $V$ we represent the supermatrix $Q$ in
the form  
\begin{equation}
Q=Vq\bar{V},\;\;\;\;V\bar{V}=1  \label{e1.15a}
\end{equation}%
where $q$ is a diagonal matrix $[q,\Lambda ]=0$. One can impose the
condition $V\Lambda =\Lambda \bar{V}$ to fix the gauge freedom related to
the invariance of the definition, Eq.(\ref{e1.15a}), with respect to the
replacement $V\rightarrow Vv$, where $[v,\Lambda ]=0$. The spectrum of
fluctuations of the matrix $q$ has a gap and they can be considered using
the saddle-point approximation. Substitution of Eq.(\ref{e1.15a}) into the
free energy functional, Eq.(\ref{e1.10}), and integration over $q$ carried
out in the quadratic in fluctuations of $q$ approximation results in the
following expression for the effective free energy functional 
\begin{equation}
F[V]=\frac{i}{2}{\rm Str}\left( \bar{V}\left[\hat{H}+\frac{\omega +i\delta }{2}%
\Lambda ,V\right] g\right) -\frac{1}{4}{\rm Str}\bigl(\bar{V}[\hat{H},V]g\bigr)%
^{2},  \label{e1.16}
\end{equation}%
where $g$ is the Green function, Eq. (\ref{e1.7}), in the absence of the
source $a=0$. This expression has been first obtained in Ref.\cite{EST04}
and is just the first two terms of the expansion of the free energy
functional in powers of the commutator $[\hat{H},V]$. As it has been shown in Ref.%
\cite{EST04}, this approximation is sufficient if the scattering is rather
weak. We use Eq.(\ref{e1.16}) as the starting point for the study of RMF
problem and consider the limit of a weak field.

\section{RMF: formulation of the problem. Reduction to collisional $\protect%
\sigma $-model.}

\label{collision}

In order to take into account the spin of the electrons one should double
the number of variables and consider in Eqs.(\ref{e1.1}), (\ref{e1.2}) $16$%
-component supervectors $\psi ({\bf r})$. The most convinient choice for
their structure is as follows (see also \cite{E83}): 
\[
\psi=\left(
\begin{array}{ccc}
\psi^1  \\
\psi^2 
\end{array}\right),\;\;\;
\psi ^{m}=\left( 
\begin{array}{ccc}
\vartheta ^{m} \\ 
v^{m} 
\end{array}%
\right) ,
\]
\begin{equation}
\vartheta^{m}=\frac{1}{\sqrt{2}}\left( 
\begin{array}{ccc}
{\chi ^{m}}^{\ast } \\ 
i\sigma _{y}\chi ^{m}  
\end{array}%
\right) ,\text{ \ \ }v^{m}=\frac{1}{\sqrt{2}}\left( 
\begin{array}{ccc}
{S^{m}}^{\ast } \\ 
i\sigma _{y}S^{m}   
\end{array}%
\right)   \label{e1.17}
\end{equation}
where $m=1,2$ divides the supervector space into advanced (A) and retarded
(R) subspaces; $\chi ^{m},S^{m}$ are anti- and commuting $2$-component
vectors in spin space, $\sigma _{y}$ is the second Pauli matrix. The matrix 
$C$ that determines the charge conjugation is written now as 
\[
C=\Lambda \otimes \left( 
\begin{array}{ccc}
c_{1} & 0 \\ 
0 & c_{2}  
\end{array}
\right) 
\]
\begin{equation}
c_{1}=\left( 
\begin{array}{ccc}
0 & i\sigma _{y} \\ 
i\sigma _{y} & 0  
\end{array}%
\right) ,\text{ \ \ }c_{2}=\left( 
\begin{array}{ccc}
0 & -i\sigma _{y} \\ 
i\sigma _{y} & 0  
\end{array}%
\right)   \label{e1.18}
\end{equation}
where the matrix $\Lambda $ is the third Pauli matrix in the
advanced-retarded space. Below we use also the matrix $\tau _{3}=\pm 1$ in
the time-reversal space and ${\bf \Sigma }=\tau _{3}\otimes {\bf \sigma }$, 
${\bf \sigma }=(\sigma _{x},\sigma _{y},\sigma _{z})$. Using these
definitions we write the Hamiltonian $\hat{H}_{{\bf r}}$, Eq.(\ref{e1.2}),
as follows 
\begin{equation}
\hat{H}_{{\bf r}}=\frac{1}{2m}\left[ -i{\bf \nabla _{r}}-\frac{e}{c}\tau _{3}%
{\bf A}({\bf r})\right] ^{2}-\epsilon _{F}-\frac{g}{2}\mu _{B}{\bf B}({\bf r}%
){\bf \Sigma },  \label{e1.19}
\end{equation}%
$\mu _{B}=e/(2mc)$, $g$ is a factor that determines the Zeeman splitting. In
Ref.\cite{ET02} the authors put $g=2$ and assumed that the magnetic field 
${\bf B}({\bf r})$ was perpendicular to the plane of electron gas. This was
the only case when the mathematical trick used in that paper and based on 
replacing the initial Hamiltonian (\ref{e1.19}) by the Dirac Hamiltonian could 
be realized. In the present paper, using
the new scheme of the calculations we consider a more general case with an
arbitrary value of $g$-factor and the direction of the magnetic field. 

As mentioned previously, we consider the limit of a weak magnetic field.
This assumption implies that the corresponding radius $R_{H}$ of the
cyclotron motion $R_{H}=v_{F}(eB/mc)^{-1}$, $v_{F}$ is Fermi velocity, is
larger than other characteristic lengths. Then, considering electrons in a
region with sizes smaller than $R_{H}$ one can choose the vector potential 
${\bf A}({\bf r})$ such that it would be small everywhere in this region. The
smallness of the vector potential allows one to represent the Hamiltonian
Eq.(\ref{e1.19}) as the sum of the main part $\hat{H}_{0}$ and a small
perturbation $\delta \hat{H}$ 
\[
\hat{H}_{{\bf r}}=\hat{H}_{0{\bf r}}+\delta \hat{H}_{{\bf r}},\text{ \ \ }%
\hat{H}_{0{\bf r}}=-\frac{{\bf \nabla _{r}}^{2}}{2m}-\epsilon _{F},\]
\begin{equation}
\delta \hat{H}_{{\bf r}}=i\frac{e}{mc}\tau _{3}{\bf A}({\bf r}){\bf \nabla
_{r}}+\frac{e^{2}}{2mc^{2}}{\bf A}^{2}({\bf r})-\frac{g}{2}\mu _{B}{\bf %
\Sigma B}({\bf r})  \label{e1.20}
\end{equation}
We choose for the vector potential the following gauge:
\[
{\rm div}_{{\bf r}}{\bf A}({\bf r},z=0)\equiv \partial _{x}A_{x}({\bf r}%
,z=0)+\partial _{y}A_{y}({\bf r},z=0)=0,\]
\[
A_{z}({\bf r},z=0)=0,
\]
${\bf r}$ is coordinate on the plane of electron gas. This gauge corresponds
to the well-known London gauge in the superconductivity theory.

The next step in the calculation is to average the partition function 
$Z[a]=\int \exp \bigl(-F[V]\bigr)DV$ with $F[V]$ from Eq.(\ref{e1.16}) over
the random magnetic field. In the present paper we take a gaussian
distribution of the field with the pair correlation 
\begin{equation}
\langle B_{i}({\bf r})B_{j}({\bf r^{\prime }})\rangle =2\left( \frac{mc}{e}%
\right) ^{2}w_{ij}({\bf r}-{\bf r^{\prime }})  \label{e1.21}
\end{equation}%
The free energy $F[V]$, Eq.(\ref{e1.16}), is written in a somewhat implicit
way. In order to reduce it to a more explicit way we use the Wigner
representation.

For any matrix $O$ it is defined as:  
\begin{equation}
O({\bf r}+{\bf \rho }/2,{\bf r^{\prime }}-{\bf \rho }/2)=\int O_{{\bf p}}(%
{\bf r})e^{i{\bf p\rho }}d{\bf p}  \label{e1.21a}
\end{equation}%
The product of  two operators $\hat{O}_{1}$, $\hat{O}_{2}$ can be written in
the Wigner representation in the following form: 
\[
O_{1{\bf p}}({\bf r})\ast Q_{2{\bf p}}({\bf r})=
\]
\begin{equation}
O_{1{\bf p}}({\bf r})\exp\left[\frac{i}{2}\left(
\overleftarrow{{\bf \nabla}}_{{\bf r}}\overrightarrow{{\bf\nabla}}_{{\bf p}}-
\overleftarrow{{\bf \nabla}}_{{\bf p}}\overrightarrow{{\bf\nabla}}_{{\bf r}}
\right)\right]
O_{2{\bf p}}({\bf r})  \label{e1.21b}
\end{equation}
The new operation $\ast $ introduced in Eq.(\ref{e1.21b}) has all properties
of the usual matrix product. In particular, it is associative.

Using the Wigner representation we expand Eq.(\ref{e1.16}) in 
$\delta\hat{H}$ and regard only terms up to the second order 
\begin{equation}
F[V]=F_{0}[V]+F_{1}[V]+F_{2}[V],  \label{e1.220}
\end{equation}
where r.h.s. is the sum of terms of the zeroth, first and second order in
the field, respectively. Below we present an explicit calculation of the
zeroth term only. Calculation of the other terms can be carried out in the
same way.

The zeroth order term $F_{0}[V]$ takes the form 
\[
F_{0}[V]=\frac{i}{2}{\rm Str}\left( \bar{V}\left[{\cal H}_{0},V\right]
g_{0}\right) -\frac{1}{4}{\rm Str}\bigl(\bar{V}[{\cal H}_{0},V]g_{0}\bigr)^{2},\] 
\begin{equation}
{\cal H}_{0}=\frac{\hat{{\bf p}}^{2}}{2m}-\epsilon _{F}+
\frac{\omega +i\delta }{2}\Lambda
\label{e1.22}
\end{equation}%
$g_{0{\bf p}}$ is the Fourrier transformed Green function Eq.(\ref{e1.7}) in the 
absence of the field, ${\bf A}=0$ and the source $a=0$. We find the commutator 
$[{\cal H}_{0},V]$ and write it in the Wigner representation as 
\begin{equation}
i[{\cal H}_{0},V]=\frac{{\bf p\nabla _{r}}}{m}V_{{\bf p}}({\bf r})+\frac{i(\omega
+i\delta )}{2}[\Lambda ,V_{{\bf p}}({\bf r})]  \label{e1.22a}
\end{equation}%
Eq.(\ref{e1.22a}) contains a small frequency $\omega $ and the gradient 
${\bf \nabla _{r}}V_{{\bf p}}({\bf r})$. Being interested in variations at distances
much exceeding the wave lentgh we neglect the second term in Eq.(\ref{e1.22}%
). In the same approximation one may replace the ``star product'' $\ast $
Eq.(\ref{e1.21b}) by the usual one and write the first term in Eq.(\ref%
{e1.22}) as follows 
\begin{equation}
F_0[V]=\frac{1}{2}\int \bar{V}_{{\bf p}}({\bf r})\left[ \frac{{\bf p\nabla _{r}}}{m}%
+\frac{i(\omega +i\delta )}{2}\Lambda \right] V_{{\bf p}}({\bf r})g_{0{\bf p}%
}d{\bf r}d{\bf p},  \label{e1.22b}
\end{equation}%
The Green function $g_{0{\bf p}}$ has a sharp peak at the Fermi surface,
whereas the function $V_{{\bf p}}$ is smooth. Therefore, one may replace in
Eq.(\ref{e1.22b}) the momentum ${\bf p}$ by its value at the surface: ${\bf p}%
\rightarrow p_{F}{\bf n}$, consider separately the integration over the
absolute value $|{\bf p}|$, which is equivalent to integration over $\xi =%
{\bf p}^{2}/2m-\epsilon _{F}$,  and unit vector ${\bf n}={\bf p}/p$. Making
the replacement $d{\bf p}\rightarrow \nu d\xi d{\bf n}$ and carrying out the
integration over $\xi $  only in the Green function $g_{0{\bf p}}$ we have 
\begin{equation}
\int g_{0{\bf p}}d\xi =i\int \left[ \xi -\epsilon +\frac{\omega }{2}%
(1+\Lambda )+i\delta \Lambda \right] ^{-1}d\xi =\pi \Lambda   \label{e1.22c}
\end{equation}%
Using Eq.(\ref{e1.22c}) we come to the following result for $F_{0}$, Eq.(\ref{e1.22b}) 
\begin{equation}
F_{0}[V]=\frac{\pi \nu }{2}{\rm Str}\int \Lambda \bar{V}_{{\bf n}}({\bf r}%
)\left(v_{F}{\bf n\nabla _{r}}+\frac{i(\omega +i\delta )}{2}\Lambda
\right) V_{{\bf n}}({\bf r})d{\bf r}d{\bf n},  \label{e1.22d}
\end{equation}%
This expression is just the usual kinetic term of the ballistic $\sigma $%
-model with the matrix $Q_{{\bf n}}=V_{{\bf n}}\Lambda \bar{V}_{{\bf n}}$.
In order to take into consideration the scattering one should calculate the
rest terms $F_{1}[V]$, $F_{2}[V]$ of Eq.(\ref{e1.220}).

In the main in small commutator $[H_{0},V]$, Eq.(\ref{e1.22a}),
approximation the term $F_{1}[V]$, Eq.(\ref{e1.220}), may be written in the
following form:
\begin{equation}
F_{1}[V]=\frac{i}{2}{\rm Str}\bigl(\bar{V}[\delta H,V]g_{0}\bigr)
\label{e1.23}
\end{equation}
Using the Wigner representation and neglecting high gradients of the
matrices $V_{{\bf p}}({\bf r})$, $\bar{V}_{{\bf p}}({\bf r})$  we find with
the help of Eq. (\ref{e1.20})  
\[
F_{1}[V]=i\frac{e}{2mc}{\rm Str}\int e^{i{\bf qr}}\left[ -\tau _{3}{\bf pA%
}({\bf q})-(g/4){\bf \Sigma B}({\bf q})\right] \times   \]
\begin{equation}
V_{{\bf p}-\frac{{\bf q}}{2}}({\bf r})g_{0{\bf p}}\bar{V}_{{\bf p}+\frac{%
{\bf q}}{2}}({\bf r})d{\bf r}d{\bf p}d{\bf q},  \label{e1.23a}
\end{equation}
In Eq.(\ref{e1.23a}), we disregard the term $(e^{2}/2mc^{2}){\bf A}^{2}({\bf %
r})$ in $\delta H_{{\bf r}}$, Eq.(\ref{e1.20}), because it results in small
in the magnetic field corrections. Moreover, considering large distances and
therefore assuming that the matrices $V_{{\bf p}}({\bf r})$ vary in space
slower than the vector potential ${\bf A}({\bf r})$ and magnetic field ${\bf %
B}({\bf r})$ we conclude that the functional $F_{1}[V]$ should be
self-averaging and vanish. Thus, only term $F_{2}[V]$ in the expansion Eq.(%
\ref{e1.220}) gives a contribution related to the scattering in the
RMF. In the main order it reads: 
\[
F_{2}[V]=-\frac{1}{2}{\rm Str}\bigl(\bar{V}[\delta H,V]g_{0}\delta Hg_{0}%
\bigr)-\frac{1}{4}{\rm Str}\bigl(\bar{V}[\delta H,V]g_{0}\bigr)^{2}\approx 
\]
\[
-\left( \frac{e}{2mc}\right) ^{2}{\rm Str}\int \left[ -\tau _{3}{\bf pA}(%
{\bf q})+(g/4){\bf \Sigma B}({\bf q})\right] \times   \]
\[V_{{\bf p}-\frac{{\bf q}}{2}}({\bf r})g_{0{\bf p}-\frac{{\bf q}}{2}}\bar{V}%
_{{\bf p}+\frac{{\bf q}}{2}}({\bf r})\left[ -\tau _{3}{\bf pA}({\bf %
q^{\prime }})+(g/4){\bf \Sigma B}({\bf q^{\prime }})\right] \times  
\]
\begin{equation}
V_{{\bf p}-\frac{{\bf q^{\prime }}}{2}}({\bf r})g_{0{\bf p}-\frac{{\bf %
q^{\prime }}}{2}}\bar{V}_{{\bf p}+\frac{{\bf q^{\prime }}}{2}}({\bf r})e^{i(%
{\bf q}+{\bf q^{\prime }}){\bf r}}d{\bf r}d{\bf p}d{\bf q}d{\bf q^{\prime }}
\label{e1.24}
\end{equation}
Eq.(\ref{e1.24}) can be easily averaged over the magnetic field. Noticing again 
that $g_{0{\bf p}}$ is a sharp function of the momentum ${\bf p}$ at the Fermi 
surface and using Eq.(\ref{e1.21}) we come to the $\sigma $ model with the following effective
energy: 
\[
F[Q_{{\bf n}}]=F_{kin}[Q_{{\bf n}}]+F_{orb}[Q_{{\bf n}}]+F_{sp}[Q_{{\bf n}%
}]+F_{sp-orb}[Q_{{\bf n}}],
\]%
where $F_{kin}[Q_{{\bf n}}]$ determines ballistic (or free) propagation and
coincides with $F_{0}[Q_{{\bf n}}]$ Eq.(\ref{e1.22d}) whereas the other
terms are related to three types of scattering which can be called as
orbital, spin and spin-orbital, respectively: 
\begin{eqnarray}
F_{orb}[Q_{{\bf n}}] &=&-\frac{(\pi \nu )^{2}}{8}{\rm Str}\int \frac{1+{\bf n%
}_{1}{\bf n}_{2}}{1-{\bf n}_{1}{\bf n}_{2}}w_{zz}({\bf n}_{1}-{\bf n}_{2}) 
\nonumber \\
&&\times Q_{{\bf n}_{1}}({\bf r})Q_{{\bf n}_{2}}({\bf r})d{\bf r}d{\bf n}%
_{1}d{\bf n}_{2},  \label{e1.25}
\end{eqnarray}%
\begin{eqnarray}
F_{sp}[Q_{{\bf n}}] &=&-\frac{(\pi \nu )^{2}}{2}{\rm Str}\int \left( \frac{g%
}{4}\right) ^{2}w_{ij}({\bf n}_{1}-{\bf n}_{2})\times   \nonumber \\
&&Q_{{\bf n}_{1}}({\bf r})\sigma _{i}Q_{{\bf n}_{2}}({\bf r})\sigma _{j}d%
{\bf r}d{\bf n}_{1}d{\bf n}_{2},  \label{e1.26}
\end{eqnarray}%
\begin{eqnarray}
F_{sp-orb}[Q_{{\bf n}}] &=&-\frac{(\pi \nu )^{2}}{2}{\rm Str}\int \left( i%
\frac{g}{4}\right) \frac{[{\bf n}_{1}\times {\bf n}_{2}]_{z}}{1-{\bf n}_{1}%
{\bf n}_{2}}w_{zj}({\bf n}_{1}-{\bf n}_{2})   \nonumber \\
&\times&\sigma _{j}Q_{{\bf n}_{1}}({\bf r})Q_{{\bf n}_{2}}({\bf r})d{\bf r}d{\bf n}%
_{1}d{\bf n}_{2},  \label{e1.27}
\end{eqnarray}
where $w_{ij}({\bf n}_1-{\bf n}_2)$ is Fourrier transformation of the 
correlation function Eq.(\ref{e1.21}) for the momentum difference 
${\bf q}$ at the Fermi surface ${\bf q}=p_F({\bf n}_1-{\bf n}_2)$ 
and the sum over $i$, $j$ is implied in Eqs.(\ref{e1.26}), (\ref{e1.27}). 
The last term, Eq.(\ref{e1.27}), corresponding to an effective spin-orbital 
scattering describes correlations between the orbital and spin scattering. The 
functional $F_{orb}[Q_{{\bf n}}]$, Eq.(\ref{e1.25}), has been previously 
obtained in Ref.\cite{ETS01}. Let us remind that 
$Q_{{\bf n}}=V_{{\bf n}}\Lambda \bar{V}_{{\bf n}}$ in Eqs.(\ref{e1.25}-\ref{e1.27}) 
is $16\times 16$ supermatrix with an additional spin structure.

The functionals, Eqs.(\ref{e1.25}-\ref{e1.27}), have a form typical for $%
\sigma $-model in the regime that can be specified as collisional. From this
point of view ${\bf n}_{1}$, ${\bf n}_{2}$ in Eqs.(\ref{e1.25}-\ref{e1.27})
can be considered as momenta of a particle before and after collision
whereas the integrands determine corresponding transition probability.
Previously we assumed that the matrices $V_{{\bf n}}({\bf r})$ vary in space
sufficiently slowly. Now we can clarify this assumption more carefully.

We notice that its formulation is directly related to the conditions under
which the collisional regime described by Eqs.(\ref{e1.25}-\ref{e1.27})
should be used. These conditions depend on the kind of disorder. If disorder
is short ranged ($\delta $-correlated) the scattering may be considered as a
collision on the length scales exceeding the Fermi wavelength $\lambda _{F}$%
. The situation becomes more complicated when the disorder is long ranged.
This case for the potential disorder was first studied in Refs.\cite{AL96}$^{,}$ 
\cite{GM02} and then, with using results of these papers and applying the
ballistic $\sigma $-model, in Ref.\cite{EK03}. The model with a long ranged
magnetic field was directly considered in Ref.\cite{EKrmf03}. It was shown that
collisional regime corresponds to lengths exceeding the Lyapunov length $%
l_{L}=v_{F}\lambda_L^{-1}$, $\lambda_L$ is the rate of divergency of the 
classical trajectories.

Summing up,  the model found in Eqs.(\ref{e1.25}-\ref{e1.27}) should be
valid for matrices $V_{{\bf n}}({\bf r})$ slowly varying over the Fermi
wavelength $\lambda _{F}$ in the case of short ranged RMF or over the
Lyapunov length $l_{L}$ in the case of a long ranged field.

\section{Diffusive model.}

\label{diffusion} In this section we study diffusive limit of the model
obtained in Eqs.(\ref{e1.25}-\ref{e1.27}) and also calculate the spin
susceptibility in this limit.

We start the reduction of the theory given by Eqs.(\ref{e1.25}-\ref{e1.27})
to the effective $\sigma $-model applicable in the diffusive limit noticing
that the functionals, Eqs.(\ref{e1.25}-\ref{e1.27}), are invariant with
respect to rotations $U({\bf r})$ 
\begin{equation}
Q_{{\bf n}}({\bf r})\rightarrow U({\bf r})Q_{{\bf n}}({\bf r})\bar{U}({\bf r}%
),\text{ \ \ }U({\bf r})\bar{U}({\bf r})=1  \label{e1.28}
\end{equation}
provided $U({\bf r})$ is the $\delta $-function in the spin space. This
means that only the charge mode related to fluctuations of the electron
density remains gapless in the diffusive limit.

The situation changes if one considers a random magnetic field with a fixed
direction ${\bf h,}$ ${\bf h}^{2}=1,$ replacing the pair function, Eq.(\ref%
{e1.21}), by the following one:
\begin{equation}
\langle B({\bf r})B({\bf r^{\prime }})\rangle=
2\left( \frac{mc}{e}\right) ^{2}w({\bf r}-{\bf r^{\prime }}),  
\label{e1.28a}
\end{equation}
where $B({\bf r})$ is the absolute value of the magnetic field. 
Function $w({\bf r}-{\bf r^{\prime }})$ being considered in $3D$ 
should be independent of the coordinate along ${\bf h}$ due to the condition 
${\rm div}_{{\bf r}}{\bf B}=0$. One can readily see that in this case the 
energies Eqs.(\ref{e1.25}-\ref{e1.27}) are invariant with respect to the 
transformation, Eq.(\ref{e1.28}), if $U({\bf r})$ commutes with the matrix 
$\sigma_{h}\equiv ({\bf \sigma h})$ and, consequently, an additional soft mode 
identified with the spin mode along ${\bf h}$ appears. Considering only 
charge and spin modes, $[Q_{{\bf n}},\sigma _{h}]=0,$ and simplifying 
Eqs.(\ref{e1.25}-\ref{e1.27}) we come to the following form of collision term 
\[
F_{scatt}[Q_{{\bf n}}]=
-\frac{(\pi \nu )^{2}}{8}{\rm Str}\int \left[ \cos ^{2}\theta \frac{1+{\bf n%
}_{1}{\bf n}_{2}}{1-{\bf n}_{1}{\bf n}_{2}}+\left( \frac{g}{2}\right) ^{2}%
\right]\times   \]
\begin{equation}
w({\bf n}_{1}-{\bf n}_{2})Q_{{\bf n}_{1}}({\bf r})Q_{{\bf n}_{2}}({\bf r})d%
{\bf r}d{\bf n}_{1}d{\bf n}_{2},  \label{e1.29}
\end{equation}
where $\cos\theta =({\bf e}_{z}{\bf h})$, ${\bf e}_z$ is $z$-ort. Because of the 
usual space symmetry the term $F_{sp-orb}[V]$, Eq.(\ref{e1.27}), vanishes and 
does not contribute to the energy Eq.(\ref{e1.29}). 

From Eq.(\ref{e1.29}) one can come to the diffusive $\sigma$ model 
singling out the angle modes $\tilde{Q}_{{\bf n}}({\bf r})$

\[
Q_{{\bf n}}({\bf r})=U({\bf r})\tilde{Q}_{{\bf n}}({\bf r})\bar{U}({\bf r}),
\]
that describe the fluctuation of the particle momentum ${\bf n}$ and
integrating them out. Calculation of integrals over the angle modes $\tilde{Q%
}_{{\bf n}}$ is standard and can be found in details e.g. in Ref.\cite{ETS01}%
. The final expression for the effective energy has the usual form  
\begin{equation}
F[Q({\bf r})]=\frac{\pi \nu }{8}{\rm Str}\int \left[ D({\bf \nabla _{r}}%
Q)^{2}+2i\omega \Lambda Q({\bf r})\right] d{\bf r}  \label{e1.30}
\end{equation}%
where $D=v_{F}^{2}\tau _{tr}/2$, $\tau _{tr}$ is the transport time 
\begin{eqnarray}
\tau _{tr}^{-1}=\pi \nu \int  &&\left[ \cos ^{2}\theta (1+{\bf n}_{1}{\bf n}%
_{2})+\left( \frac{g}{2}\right) ^{2}(1-{\bf n}_{1}{\bf n}_{2})\right]  
\nonumber \\
&&\times w({\bf n}_{1}-{\bf n}_{2})d{\bf n}_{1}d{\bf n}_{2}  \label{e1.31}
\end{eqnarray}%
This time determines the relaxation of the angle modes and, hence, of the
particle momentum ${\bf n}$.

If the magnetic field is $\delta $-correlated in space the transport time
can be explicitly calculated and expressed through the time $\tau
_{tr}^{\prime }$ in the model of spinless electrons and perpendicular 
magnetic field: 
\begin{equation}
\tau _{tr}^{-1}={\tau _{tr}^{\prime }}^{-1}\left[ \cos ^{2}\theta +(g/2)^{2}%
\right]   \label{e1.32}
\end{equation}%
In the case of the magnetic field perpendicular to the plane, $\cos \theta
=1,$ and free electrons, $g=2$, the transport time $\tau _{tr}$ turns out
two times smaller than the transport time for the spinless electrons. The
same result for this particular cas has been established in Ref.\cite{ET02}.

Using Eqs.(\ref{e1.25}-\ref{e1.27}), (\ref{e1.30}) one can calculate various
physical quantities. Below we find the spin magnetic susceptibility reliing 
on the diffusive model Eq.(\ref{e1.30}).

First, we use the Kubo formula and write the magnetic moment due to the spin
of electrons as a sum of two terms: 
\[
{\bf m}({\bf r},\omega )={\bf m}_{sp}({\bf r},\omega )+{\bf m}_{so}({\bf r}%
,\omega )
\]%
where ${\bf m}_{sp}$ is a purely spin contribution:  
\[
{\bf m}_{sp}({\bf r},\omega )=i\mu _{B}^{2}\frac{g}{2}\int_{-\infty
}^{+\infty }\frac{d\epsilon }{2\pi }\left[ n(\epsilon -\omega )-n(\epsilon )%
\right] \times\]
\[
{\rm Tr}\int \left[ G_{\epsilon }^{R}({\bf r},{\bf r^{\prime }})({\bf %
\sigma }\cdot {\bf B}({\bf r^{\prime }},\omega ))G_{\epsilon -\omega }^{A}(%
{\bf r^{\prime }},{\bf r}){\bf \sigma }\right] d{\bf r^{\prime }}+  \]
\[
i\mu _{B}^{2}\frac{g}{2}\int_{-\infty }^{+\infty }n(\epsilon ){\rm Tr}\int %
\left[ G_{\epsilon }^{A}({\bf r},{\bf r^{\prime }})({\bf \sigma }\cdot {\bf B%
}({\bf r^{\prime }},\omega ))G_{\epsilon -\omega }^{A}({\bf r^{\prime }},%
{\bf r}){\bf \sigma }-\right. \]
\begin{equation}
\left. G_{\epsilon +\omega }^{R}({\bf r},{\bf r^{\prime }}){\bf \sigma }%
\cdot {\bf B}({\bf r^{\prime }},\omega ))G_{\epsilon }^{R}({\bf r^{\prime }},%
{\bf r}){\bf \sigma }\right] d{\bf r^{\prime }},  \label{e1.33}
\end{equation}
${\rm Tr}$ corresponds to the trace in the usual spin space. The other term $%
{\bf m}_{so}$ is given by the same formula taken after the replacement $(g/4)%
{\bf \sigma B}({\bf r^{\prime }},\omega )\rightarrow -i{\bf A}({\bf %
r^{\prime }},\omega ){\bf \nabla _{r}^{\prime }}$. This contribution is
related to the both orbital and spin motions. It is determined by the angle
modes and interaction between the charge and spin modes. Both factors are
irrelevant in diffusive limit and the contribution is small.

So, we neglect ${\bf m}_{so}$ and calculate ${\bf m}_{sp}$. The contribution
to this quantity given by the second term in Eq.(\ref{e1.33}) can be easily
found and corresponds to the usual Pauli susceptibility. In order to
calculate the first term we express the product of two Green functions $G^{R}
$ and $G^{A}$ through the partition function, Eqs.(\ref{e1.1}), (\ref{e1.9}%
): 
\begin{equation}
{\rm Tr}\left[ G_{\epsilon }^{R}({\bf r}_{1},{\bf r}_{2})\sigma
_{j}G_{\epsilon -\omega }^{A}({\bf r}_{2},{\bf r}_{1})\sigma _{i}\right]
=\left. -\frac{\partial ^{2}Z[a]}{\partial \alpha _{1}\partial \alpha _{2}}%
\right| _{\alpha _{1},\alpha _{2}=0},  \label{e1.34}
\end{equation}%
where the source $a({\bf r},{\bf r^{\prime }})$ is taken as follows: 
\[
a({\bf r},{\bf r^{\prime }})=\left( 
\begin{array}{ccc}
0 & a_{12}({\bf r}) &  \\ 
a_{21}({\bf r}) & 0 & 
\end{array}%
\right) \delta ({\bf r}-{\bf r^{\prime }})
\]%
\begin{equation}
a_{12}({\bf r})=\frac{1-k}{2}\otimes \left( 
\begin{array}{ccc}
-\alpha _{2}\sigma_{j}^T\delta ({\bf r}-{\bf r}_{2}) & 0 \\ 
0 & \alpha _{1}\sigma _{i}^T\delta ({\bf r}-{\bf r}_{1})  
\end{array}%
\right)   \label{e1.35}
\end{equation}%
\[
a_{21}({\bf r})=\frac{1-k}{2}\otimes \left( 
\begin{array}{ccc}
\alpha _{1}\sigma_{i}^T\delta ({\bf r}-{\bf r}_{1}) & 0 \\ 
0 & -\alpha _{2}\sigma_{j}^T\delta ({\bf r}-{\bf r}_{2})  
\end{array}%
\right) 
\]%
Substitution of the partition function $Z[a]$ taken in diffusive limit into
Eq.(\ref{e1.34}) results in the following expression for the susceptibility: 
\[
\chi _{ij}({\bf q},\omega )=g\mu _{B}^{2}\nu \biggl[\delta _{ij}-i\omega 
\frac{\pi \nu }{16}\langle {\rm Tr}\biggl(\sigma_{i}^T\bigl[Q_{84}({\bf q}%
)+Q_{37}({\bf q})\bigr]\biggr)\times 
\]%
\begin{equation}
{\rm Tr}\biggl(\sigma_{j}^T\bigl[Q_{73}(-{\bf q})+Q_{48}(-{\bf q})\bigr]%
\biggr)\rangle _{Q}\biggr]  \label{e1.36}
\end{equation}%
The first term in the brackets is the usual Pauli susceptibility. The other
term determines the correction to this result related to RMF. The angular
brackets $\langle \dots \rangle _{Q}$ in Eq.(\ref{e1.36}) imply the
averaging of the quantity inside them over the matrix $Q$ with the free
energy functional, Eq.(\ref{e1.30}).

We calculate this average using the parametrization 
\begin{equation}
Q({\bf r})=\Lambda \bigl[1+iP({\bf r})\bigr]\bigl[1-iP({\bf r})\bigr]^{-1}
\label{e1.37}
\end{equation}%
where the non-diagonal matrix $P({\bf r})$, $\bar{P}({\bf r})=-P({\bf r})$, $%
[P,\Lambda ]=0$ is the sum of the charge and spin modes 
\begin{equation}
P({\bf r})=P_{ch}({\bf r})+({\bf \Sigma h})P_{sp}({\bf r})  \label{e1.38}
\end{equation}
$P_{ch}({\bf r})$, $P_{sp}({\bf r})$ are $\delta$-functions in the spin 
space. Their structure should be found in correspondence with Eq.(\ref{e1.10a}) 
and condition $[Q,\hat{\tau}_3]=0$. This results in the following 
representation:
\[
P_{ch(sp)}({\bf r})=\left(
\begin{array}{ccc}
0 & B_{ch(sp)}({\bf r}) \\
\bar{B}_{ch(sp)}({\bf r}) & 0
\end{array}\right),\]
\begin{equation}
B_{ch(sp)}({\bf r})=\left(
\begin{array}{ccc}
\hat{a}_{ch(sp)}({\bf r}) & i\hat{\rho}_{1ch(sp)}({\bf r}) \\
\hat{\rho}_{2ch(sp)}^+({\bf r}) & i\hat{b}_{ch(sp)}({\bf r})
\end{array}\right),
\label{e1.38a}\end{equation}
where $\bar{B}_{ch(sp)}({\bf r})=C_0B_{ch(sp)}^T({\bf r})C_0^T$, 
$\hat{a}_{ch(sp)}({\bf r})$, $\hat{b}_{ch(sp)}({\bf r})$ are usual and 
$\hat{\rho}_{1,2ch(sp)}$ -- Grassman matrices 
$4\times4$ unit, in spin and particle-hole spaces:
\[
\hat{a}_{ch(sp)}({\bf r})=\left(
\begin{array}{ccc}
a_{ch(sp)}({\bf r}) & 0 \\
0 & a_{ch(sp)}^*({\bf r})
\end{array}\right),\]
\[
\hat{b}_{ch(sp)}({\bf r})=\left(
\begin{array}{ccc}
b_{ch}(sp)({\bf r}) & 0 \\
0 & b_{ch(sp)}^*({\bf r})
\end{array}\right)\]
\begin{equation}
\hat{\rho}_{1,2ch(sp)}({\bf r})=\left(
\begin{array}{ccc}
\rho_{1,2ch(sp)}({\bf r}) & 0 \\
0 & -\rho_{1,2ch(sp)}^*({\bf r})
\end{array}\right)
\label{e1.38b}\end{equation}

In Gauss approximation the charge and spin modes give independent contributions 
to the energy Eq.(\ref{e1.30}). Standard calculation of the average in 
Eq.(\ref{e1.36}) in the same approximation gives: 
\begin{equation}
\chi _{ij}({\bf q},\omega )=g\mu _{B}^{2}\nu (\delta _{ij}-h_{i}h_{j})+g\mu
_{B}^{2}\nu \frac{D{\bf q}^{2}}{D{\bf q}^{2}-i\omega }h_{i}h_{j}
\label{e1.39}
\end{equation}%
In the particular case when RMF has only the perpendicular component, this
expression is in agreement with the result of Ref\cite{ET02}.

\section{Discussion}

\label{discussion}

In the present work we considered a two-dimensional electron gas placed in a
nonuniform (random) magnetic field (RMF) using method of superbosonization
Ref.\cite{EST04}. The problem was studied in a quite general formulation
with taking into account interaction of the magnetic field with the spin of
electrons. The direction of the magnetic field was assumed random in space
and the value of $g$-factor - arbitrary. 

In the present paper we used a new method proposed in Ref.\cite{EST04}. This
method is based on the exact transformation of the initial field theory
written in terms of a functional integral over supervectors $\psi $ to a
theory described by a functional integral over supermatrices $Q$ and, by the
analogy with quantum field theory, can be called superbosonization. As the
method is exact it enabled us to deal with both the short and long range
disorder. 

The found theory turned out to be invariant with respect to 
the rotations in the superspace $Q\rightarrow VQ\bar{V}$, $V$ are 
unitary supermatrices $V\bar{V}=1$, provided that they commute with 
Hamiltonian: $[\hat{H},V]=0$. This means that the low energy limit 
of the theory is described by the rotations $V$ with a small commutator 
$[\hat{H},V]$. Conditions which should be imposed on the supermatrices 
$V$ to provide the smallness of the commutator depended on the character 
of the terms contained in Hamiltonian $H$, e.g. on the kind of disorder. 
Singling out the massive modes related to the fluctuations of the advanced 
and retarded blocks $q$ in the representation $Q=Vq\bar{V}$ where 
$q$ - arbitrary diagonal supermatrices $[q,\Lambda]=0$ and integrating 
them out we came to an effective energy functional in terms of the rotations 
$V$ only and obtained some nonlinear $\sigma$ model.

The derivation of the $\sigma $- model was carried out for an arbitrary
Hamiltonian $H$ and did not depend on its explicit form. Particularly, it
might be used for both short and long range disorder. Explicit form of the
model as well as the limits of its applicability depended on character of
terms contained in Hamiltonian $H$. In this paper we dealt only with a weak
RMF without an additional potential disorder and considered the theory in
the regime that may be called collisional. This regime was achieved
differently depending on correlation length of RMF. If the field was $\delta 
$-correlated the model was valid at distances larger than wavelength $%
\lambda _{F}$. In the case of long range RMF they should be larger than the
Lyapunov length $l_{L}$. 

After averaging the free energy functional over RMF we obtained the $\sigma $%
- model with a collision term consisting of three contributions that could
be called orbital, spin and spin-orbital. The first contribution came from
the part of the Hamiltonian acting on the orbital motion, the second one-
from the part acting on spin, whereas the last one came in the result of
joint averaging of them and describes correlations between the spin and orbit
scattering. 

The collisional term turned out to be invariant with respect to the rotation 
$Q_{{\bf n}}\rightarrow UQ\bar{U}$ provided that $U$ was $\delta $-function
in the spin space. This allowed us to conclude that in general only the mode
related to the charge transport or charge mode should survive in the
diffusive limit. At the same time, we have shown that in the particular case
of RMF with a fixed direction ${\bf h}$ one more spin mode along should be
regarded as well. We obtained the $\sigma $-model in the standard diffusive
form and found the transport time $\tau _{tr}$. Due to the usual space
symmetry the spin-orbital collision term vanished in the model with a not
fluctuating fixed RMF and did not contribute to $\tau _{tr}$.

Finally, in diffusive limit, we calculated the spin magnetic susceptibility.
The part of the susceptibility transversal to the direction ${\bf h}$ is
given by the usual Pauli expression whereas the longitudinal one has an
additional contribution proportional to the diffusion propagator.

We acknowledge a support from SFB 491 and SFB/Transregio 12.

\widetext


\begin{references}
\bibitem{GBG92} A. K. Geim, S. J. Bending, I. V. Grigorieva, Phys. Rev.
Lett. {\bf 69}, 2252 (1992).

\bibitem{S94} A. Smith, R. Taboryski, L. T. Hansen, C. B. S\o rensen, Per
Hedeg\aa rd and P. E. Lindelof, Phys. Rev. B {\bf 50}, 14726 (1994).

\bibitem{M95} F. B. Mancoff, R. M. Clarke, C. M. Marcus, S. C. Zhang, K.
Campman, A. C. Gossard, Phys. Rev. B {\bf 51}, 13269 (1995).

\bibitem{HLR93} V. Kalmeyer and S.-C. Zhang, Phys. Rev. B {\bf 46}, 9889
(1992); B. I. Halperin, P. A. Lee and N. Read, Phys. Rev. B {\bf 47},
7312(1993).

\bibitem{IL89} L. B. Ioffe and A. I. Larkin, Phys. Rev. B, {\bf 39}, 8988
(1989); N. Nagaosa and P. A. Lee, Phys. Rev. Lett. {\bf 64}, 2450 (1990);
Phys. Rev. B {\bf 45}, 966 (1992).

\bibitem{SN93} T. Sugiyama and N. Nagaosa, Phys. Rev. Lett. {\bf 70},1980
(1993).

\bibitem{LC94} D. K. K. Lee and J. T. Chalker, Phys. Rev. Lett. {\bf 72},
1510 (1994).

\bibitem{BSK98} M. Batsch, L. Schweitzer and B. Kramer, Physica B, {\bf 251}%
, 792 (1998).

\bibitem{KWAZ93} V. Kalmeyer, D. Wei, D. P. Arovas and S. Zhang, Phys. Rev.
B {\bf 48}, 11095 (1993).

\bibitem{AHK93} Y. Avishai, Y. Hatsugai and M. Kohmoto, Phys. Rev. B {\bf 47}%
, 9561 (1993).

\bibitem{KO95} T. Kawarabayashi and T. Ohtsuki, Phys. Rev. B {\bf 51}, 10897
(1995).

\bibitem{LXSZ95} D. Z. Liu, X. C. Xie, S. Das Sarma and S. C. Zhang, Phys.
Rev. B {\bf 52}, 5858 (1995); X. C. Xie, X. R. Wang and D. Z. Liu, Phys.
Rev. Lett. {\bf 80}, 3563 (1998).

\bibitem{YB97} K. Yang and R. N. Bhatt, Phys. Rev. B {\bf 55}, R1922 (1997).

\bibitem{SW00} D. N. Sheng and Z. Y. Weng, Europhys. Lett. {\bf 50}, 776
(2000).

\bibitem{MW96} J. Miller and J. Wang, Phys. Rev. Lett. {\bf 76}, 1461 (1996).

\bibitem{F99} A. Furusaki, Phys. Rev. Lett. {\bf 82}, 604 (1999).

\bibitem{C01} V. Z. Cerovski, Phys. Rev. B {\bf 64}, 161101R (2001).

\bibitem{AMW94} A. G. Aronov, A. D. Mirlin and P. W\"olfle, Phys. Rev. B 
{\bf 49}, 16609 (1994).

\bibitem{TSE00} D. Taras-Semchuk and K. B. Efetov, Phys. Rev. Lett. {\bf 85}
1060 (2000); also A. D. Mirlin and P. W\"olfle, {\it ibid.} {\bf 86}, 3688
(2001); D. Taras-Semchuk and K. B. Efetov, {\it ibid.} {\bf 86}, 3689 (2001).

\bibitem{ETS01} D. Taras-Semchuk, K.B. Efetov, Phys. Rev. B {\bf 64}, 115301
(2001).

\bibitem{EKrmf03} K. B. Efetov and V. R. Kogan, Phys. Rev. B {\bf 68},
245313 (2003).

\bibitem{E83} K. B. Efetov, Adv. Phys. {\bf 32}, 53 (1983); {\it %
Supersymmetry in Disorder and Chaos} (Cambridge Univ. Press, 1997).

\bibitem{ET02} K. Takahashi and K. B. Efetov, Phys. Rev. B {\bf 66}, 165304
(2002).

\bibitem{EST04} K. B. Efetov, G. Schwiete, K. Takahashi, Phys. Rev. Lett. 
{\bf 92}, 026807 (2004).

\bibitem{EK03} K.B. Efetov, V. R. Kogan, Phys. Rev. B {\bf 67}, 245312
(2003).

\bibitem{AL96} I.L. Aleiner, A.I. Larkin, Phys. Rev. B{\bf 54}, 14423 (1996).

\bibitem{GM02} I.V. Gornyi, A.D. Mirlin, J. Low Temp. Phys. {\bf 126} (3-4),
1339, (2002).
\end{references}
\end{document}